\documentclass[aps, reprint]{revtex4-2}
\usepackage{amsfonts}
\usepackage{amsmath}
\usepackage{amssymb}
\usepackage{graphicx}
\usepackage{braket}
\usepackage{bbm}
\usepackage{microtype} 

\usepackage[justification=Justified]{caption}
\begin{document}
\title{Spontaneous symmetry breaking for nonautonomous pseudo-Hermitian systems}
\author{L. F. Alves da Silva and M. H. Y. Moussa}
\affiliation{Instituto de Física de São Carlos, Universidade de São Paulo, P.O. Box 369, São Carlos, 13560-970, SP, Brazil}

\begin{abstract}
Here we first present an alternative formulation of the Lewis \& Riesenfeld theorem for solving the Schrödinger equation with nonautonomous Hermitian and pseudo-Hermitian Hamiltonians. We then employ this framework to characterize the spontaneous breaking of time-dependent antilinear symmetries of these Hamiltonians. We demonstrate that, under unbroken antilinear symmetries, the Lewis \& Riesenfeld phases are real and odd functions of time, which allows us to recover the well-known real spectra of time-independent pseudo-Hermitian Hamiltonians. However, in the spontaneously broken regime, imaginary components of the Lewis \& Riesenfeld phases arise, leading to coalescence effects analogous to those in the time-independent scenario. Finally, we present an illustrative example of unbroken and broken $\mathcal{PT}$-symmetry for a time-dependent Hamiltonian modeling the non-Hermitian dynamical Casimir effect.
\end{abstract}

\maketitle
\newpage

The existence of real spectra for non-Hermitian Hamiltonians exhibiting antilinear symmetries \cite{Bender98}, together with a new metric to ensure norm conservation \cite{Mostafa2002}, represents the triumph of non-Hermitian quantum mechanics as a consistent theory. $\mathcal{PT}$-symmetric systems, which are invariant under the combination of parity ($\mathcal{P}$) and time-reversal ($\mathcal{T}$), have been a typical example of this breakthrough since the seminal paper by Bender and Boettcher \cite{Bender98}. They showed that a $\mathcal{PT}$-symmetric autonomous Hamiltonian $H$ exhibits a real spectrum, despite its non-Hermiticity. However, when this symmetry is spontaneously broken, the spectrum becomes partially real and partially complex  \cite{Bender99}. In these autonomous systems, $\mathcal{PT}$-symmetry is unbroken when each eigenstate $\psi$ of the Hamiltonian ($H\psi=\varepsilon\psi$) is simultaneously an eigenstate of $\mathcal{PT}$ ($\mathcal{PT}\psi=\psi$), despite the antilinearity of the latter, such that $\varepsilon^{\ast}=\varepsilon$. The spontaneous symmetry breaking (SSB) occurs when $H$ and $\mathcal{PT}$ cease to share the same eigenstate basis, leading to the emergence of the exceptional points where the real eigenvalues split into a pair of complex conjugate values, $\varepsilon$ for $\psi$ and $\varepsilon^\ast$ for $\mathcal{PT}\psi$. A range of dramatic effects has been reported near the exceptional points, especially in optical systems with loss and gain \cite{gain-loss}, such as loss-induced transparency \cite{LossT}, unidirectional invisibility \cite{UI}, and coherent perfect laser absorbers \cite{CPA}. Our goal here is to present a method for determining the exceptional points of time-dependent (TD) Hamiltonians exhibiting general symmetries.

The subject of TD $\mathcal{PT}$-symmetric systems has also been investigated \cite{MostafahTD,Znojil,GW,FM, FM2,Fring-Meding,Maa2, Maa1,Ponte,Rod,Maa4,BeyondPT,TD-dilation,Cius,Fring-Real}, and intriguing phenomena have been reported such as infinite squeezing of the radiation states at a finite time \cite{Ponte} and the enhancement of Casimir photon creation \cite{Cius}. In particular, in Refs. \cite{Fring-Meding,Fring-Real} the authors have considered examples of TD pseudo-Hermitian Hamiltonians in the unbroken and broken $\mathcal{PT}$-regimes by adiabatically introducing an explicit time dependence into the parameters of an autonomous system. Starting from an instantaneous eigenvalue equation for the TD pseudo-Hermitian Hamiltonian, they demonstrate that the eigenenergies are real functions of time in the unbroken regime, whereas in the broken regime they are complex. However, a general framework for symmetry breaking of TD pseudo-Hermitian systems, beyond the adiabatic regime, has remained an open question, which we address here.

Recently, we have proposed a method for the derivation of a general TD antilinear symmetry $\mathcal{I}(t)$ for a TD Hermitian or non-Hermitian Hamiltonian $H(t)$ \cite{BeyondPT}. As an antilinear operator, $\mathcal{I}(t)$ complex-conjugates scalar coefficients, thus mapping $i\rightarrow -i$. Then, by letting $\mathcal{I}(t)$ act on both sides of the Schrödinger equation $\left[ H(t)-i\hbar \partial _{t}\right] \ket{\psi(t)} =0$, and then replacing $t$ by $-t$, we found that the transformed state vector $\ket{\tilde{\psi}(t)} =\mathcal{I}(-t)\ket{\psi(-t)} $ obeys exactly the same Schrödinger equation, governed by $H(t)$, with the original state vector $ \ket{\psi(t)} $ if the TD symmetry condition is defined as
\begin{equation}
i\hbar \partial_{t}\mathcal{I}(t) +H(-t)\mathcal{I}(t)-\mathcal{I}(t)H(t)=0.
\label{IA}
\end{equation}
For the case of time-independent (TI) symmetries, such as $\mathcal{PT}$, the antilinear symmetry condition (\ref{IA}) reduces to $\mathcal{I}H(t) \mathcal{I}^{-1}=H(-t)$ and, consequently, to the well-known commutation $\left[ H,\mathcal{I} \right] =0$ for TI Hamiltonians. It is therefore clear that a TI symmetry operator can bring about unwanted constraints on the TD parameters of the Hamiltonian, as it imposes more restrictive forms on Eq. (\ref{IA}). A general antilinear symmetry $\mathcal{I}(t)$ can be viewed as a gauged $\mathcal{PT}$-symmetry describing a reflection plane with time-varying position \cite{UZR-gPT}, or as rotations in phase space combined with time reversal \cite{BeyondPT}. When considering a linear TD symmetry $I(t)$ instead of the antilinear $\mathcal{I}(t)$, we obtain a symmetry condition that is the Lewis \& Riesenfeld (LR) definition of a dynamical invariant \cite{LR},
\begin{equation}
i\hbar\partial _{t}I(t)+\left[ I(t),H(t)\right] =0.
\label{IL}
\end{equation}

Here we present a framework to describe the unbroken and spontaneously broken regimes of TD antilinear symmetries satisfying Eq. (\ref{IA}). The central element of our discussion is the introduction of the Schrödinger operator $\mathcal{L}(t)=H(t)-i\hbar \partial _{t}$, which enters the action functional of quantum theory \cite{ActInt}. Going beyond Ref.~\cite{BeyondPT}, which established the TD antilinear symmetry condition, Eq.~(\ref{IA}), the present work develops a Schrödinger-operator-based criterion to determine the conditions under which such symmetries are unbroken or spontaneously broken. More specifically, we first introduce a generalized LR theorem for pseudo-Hermitian Hamiltonians reformulated in terms of the operator $\mathcal{L}(t)$. We then verify that there exists an eigenvalue equation for $\mathcal{L}(t)$, which enables an alternative approach to the LR invariant for solving the Schrödinger equation, and a method to describe the unbroken and spontaneously broken regimes of TD symmetries in TD Hamiltonians. Our method recovers the distinctive features of the unbroken and broken TI symmetries in TI Hamiltonians.

\textit{The Schrödinger operator}. The generalized LR theorem for pseudo-Hermitian Hamiltonians \cite{LR-NH}, as done in the Supplementary Material (SM), states that the exact solution of the Schrödinger equation $\mathcal{L}(t)\ket{\psi(t)} =0$ is given by
\begin{equation}
\ket{\psi(t)} =\sum_{n}c_{n}e^{-\frac{i}{\hbar }\alpha
_{n}(t)}\ket{n,t} ,  \label{SEsol}
\end{equation}%
expanded in the eigenstates $\ket{n,t}$ of a linear invariant $I(t)$ related to $H(t)$, which are assumed to be orthonormal with respect to a TD positive-definite metric $\rho (t)$, i.e. $\braket{m,t|n,t}_{\rho(t)}=\braket{m,t|\rho(t)|n,t}=\delta_{m,n}$. The coefficients $c_{n}$ follow from an arbitrary initial state and the well-known LR phases $\alpha _{n}(t)$ are derived from
\begin{equation}
\dot{\alpha}_{n}(t)\delta _{m,n}=\braket{m,t|\mathcal{L}(t)|n,t}_{\rho(t)}, \label{LRp}
\end{equation}
with the notation $\langle \phi|\bullet|\psi\rangle_{\rho(t)}\equiv
\langle \phi|\rho(t)\bullet|\psi\rangle$.
This equation shows that the eigenvectors of the dynamical invariant also diagonalize the Schrödinger operator $\mathcal{L}(t)$ such that 
\begin{equation}
\mathcal{L}(t)\ket{n,t}=\dot{\alpha}_{n}(t)\ket{n,t}.  \label{EEL}
\end{equation}

\textit{An alternative approach to the LR invariant.} Although the eigenvalue equation (\ref{EEL}) may be understood as a consequence of the generalized LR theorem, we stress that we can directly assume the existence of eigenvectors and eigenvalues of $\mathcal{L}(t)$ without resorting to a LR dynamical invariant. As a matter of fact, in the same way that the LR theorem requires the existence of the eigenvalue equation for the dynamical invariant $I(t)$, we can assume the possibility of constructing a similarity transformation through a unitary or nonunitary operator $S(t)$, which leads to a diagonal operator $S^{-1}(t)\mathcal{L}(t)S(t)$ with TD eigenvalues $\varepsilon_{n}(t)$ and TI eigenvectors $\ket{n}$,
\begin{equation}
S^{-1}(t)\mathcal{L}(t)S(t)\ket{n} =\varepsilon_{n}(t)\ket{n}. \label{EEL-S}
\end{equation}
We thus obtain the eigenvectors $\ket{n,t}=S(t)\ket{n}$ of $\mathcal{L}(t)$ and, through the isospectral connection between $\mathcal{L}(t)$ and $S^{-1}(t)\mathcal{L}(t)S(t)$, the eigenvalues $\varepsilon_{n}(t)$. While a Hermitian Hamiltonian demands a unitary transformation $S(t)$, the norm-conservation condition for a pseudo-Hermitian $H(t)$ demands a pseudo-unitary transformation, $S^{\dagger}(t)\rho(t) S(t) = \rho(t)$, as demonstrated in the SM. This alternative approach to the LR theorem enables us to derive the general solution of the Schrödinger equation from the eigenvalue equation $\mathcal{L}(t)\ket{n,t}=\varepsilon_{n}(t)\ket{n,t}$, as 
\begin{equation}
\ket{\psi(t)} =\sum_{n}c_{n}e^{-\frac{i}{\hbar} \int_{0}^{t}\varepsilon_{n}(\tau )d\tau } \ket{n,t},  \label{SEsola}
\end{equation}
where, by comparing Eq. (\ref{SEsola}) with Eq. (\ref{SEsol}), it follows that $\varepsilon_{n}(t) = \dot{\alpha}_n(t)$, as expected from Eq. (\ref{EEL}). The eigenvalue problem of $\mathcal{L}(t)$ provides the key tool for expressing the spontaneous breaking of TD antilinear symmetries within this new approach to the Lewis--Riesenfeld method.

Based on the above results, from now on we shift our focus from the TD Hamiltonian $H(t)$ to the Schrödinger operator $\mathcal{L}(t)$ from which we rewrite the symmetry condition in Eq. (\ref{IA}) in the form (SM)
\begin{equation}
\mathcal{I}(t)\mathcal{L}(t)\mathcal{I}^{-1}(t)=\mathcal{L}(-t).  \label{IAL}
\end{equation}

\textit{Pseudo-Hermiticity of} $\mathcal{L}(t)$. Following the reasoning in Ref. \cite{FM}, we assume that the TD Hamiltonian $H(t)$ is $\rho(t)$-pseudo-Hermitian satisfying (SM)
\begin{equation}
H^{\dagger }(t) \rho (t) - \rho(t) H(t) = i\hbar \partial_{t} \rho(t),  \label{TDPH}
\end{equation}%
and is thus a nonobservable operator \cite{MostafahTD,FM}, while all other observables $O(t)$ are pseudo-Hermitian in the usual way: $\rho(t)O(t)=O^{\dagger }(t)\rho (t)$. From Eq. (\ref{TDPH}) it follows that $\rho (t)\mathcal{L}(t) =\left[ H^{\dagger}(t)-i\hbar \partial _{t}\right] \rho (t)$ and, since the time derivative is an anti-Hermitian operator, we deduce that the Schrödinger operator $\mathcal{L}(t)$ is a pseudo-Hermitian observable,
\begin{equation}
\rho (t)\mathcal{L}(t)=\mathcal{L}^{\dagger }(t)\rho (t).  \label{PHL}
\end{equation}
In fact, considering (\ref{PHL}), we verify that the instantaneous eigenvalues of $\mathcal{L}(t)$, defined by Eq. (\ref{LRp}), are real functions of time, $\dot{\alpha}_{n}^{\ast }(t) =\dot{\alpha}_{n}(t)$, showing that while the observability of the pseudo-Hermitian $H(t)$ is lost, the same does not apply to the Schrödinger operator $\mathcal{L}(t)$ in the nonautonomous scenario. The eigenvalues of $\mathcal{L}(t)$ represent energy levels defined by the time rate of change of the dynamical $\alpha _{n}^{d}(t)$ plus the geometrical $\alpha _{n}^{g}(t)$ phases, with $\dot{\alpha}_{n}^{d}(t)= \braket{m,t|H(t)|n,t}_{\rho (t)}$ and $\dot{\alpha}_{n}^{g}(t)=-i\hbar \braket{m,t|\partial_{t}|n,t}_{\rho (t)}$ \cite{Salomon}. Therefore, instead of characterizing the SSB through the eigenvalues of the autonomous Hamiltonian, as in \cite{Bender99}, in our TD scenario, this characterization must be done through the eigenvalues of $\mathcal{L}(t)$.

\textit{Unbroken and broken symmetries}. Considering a general TD Hamiltonian $H(t)$ which can be Hermitian, non-Hermitian or pseudo-Hermitian according to Eq. (\ref{TDPH}), it is straightforward to show that there are two a priori distinct sets of eigenvectors of $\mathcal{L}(t)$: the state basis $\left\{\ket{n,t}\right\} $, as defined by Eq. (\ref{EEL}), and the transformed basis $\left\{ \mathcal{I}(-t)\ket{n,-t} \right\} $. The latter follows from multiplying Eq. (\ref{EEL}) on the left by $\mathcal{I}(t)$ and using Eq. (\ref{IAL}) to obtain
\begin{equation}
\mathcal{L}(t)\left[ \mathcal{I}(-t)\ket{n,-t} \right] =-
\dot{\alpha}_{n}^{\ast}(-t)\left[ \mathcal{I}(-t)\ket{n,-t} \right].  \label{EELs}
\end{equation}

Now, by comparing Eqs. (\ref{EEL}) and (\ref{EELs}), we are led to conjecture that the general antilinear symmetry is unbroken when $\mathcal{I}(t)$ shares the same eigenvector with $\mathcal{L}(t)$, but time-reversed, i.e.
\begin{equation}
\mathcal{I}(t)\ket{n,t} =\lambda_{n}\ket{n,-t}.  \label{13}
\end{equation}%
The eigenvalues $\lambda_{n}$ are assumed TI, so that $\ket{\tilde{\psi}(t)}$ and $\ket{\psi(t)}$ remain equivalent solutions of the Schrödinger equation, as expected in the unbroken symmetry regime \cite{Beekman}. A similar conjecture was made by Bender, Boettcher, and Meisinger \cite{Bender99}, in the TI scenario where $\left[ H,\mathcal{PT}\right]=0$, assigning the unbroken $\mathcal{PT}$-symmetric regime to the case where $H$ and $\mathcal{PT}$ share a common state basis despite $\mathcal{PT}$ being antilinear. When $H$ and $\mathcal{PT}$ cease to share the same basis, SSB takes place.

Assuming the unbroken symmetry condition through the validity of Eq. (\ref{13}) ---which is verified below for an illustrative example---, we can assert from Eq. (\ref{EELs}) that the LR phases satisfy
\begin{equation}
\alpha _{n}^{\ast }(-t)=-\alpha _{n}(t),  \label{14}
\end{equation}%
the real and imaginary parts being odd and even functions of time. For the particular case of pseudo-Hermitian systems, the phases $\alpha_{n}(t)$ are real, as follows from Eq. (\ref{PHL}), and therefore odd  functions of time. On the other hand, the general antilinear symmetry is spontaneously broken when Eq. (\ref{13}) is no longer satisfied, resulting in time-reversed complex-conjugate pairs of the eigenvalues of $\mathcal{L}(t)$, $\dot{\alpha}_{n}(t)$ for $\ket{n,t}$ and $-\dot{\alpha}^\ast_{n}(-t)$ for $\mathcal{I}(-t)\ket{n,-t}$, associated with two different solutions of the Schrödinger equation, $\ket{\psi(t)}$ and $\ket{\tilde{\psi}(t)}$, respectively. For pseudo-Hermitian systems, Eq. (\ref{14}) describes the time reversibility of the energy levels, $\varepsilon_{n}(-t)=\varepsilon_{n}(t)=\dot{\alpha}_{n}(t)$, while imaginary parts resulting from SSB provide irreversible amplification or dissipation processes, as seen from Eq. (\ref{SEsola}). It is worth noting that, unlike in the autonomous scenario, unbroken TD antilinear symmetry and TD pseudo-Hermiticity are not equivalent notions: the former constrains the time-reversal properties of the LR phases, whereas the latter enforces the instantaneous reality of the eigenvalues of $\mathcal{L}(t)$.

\textit{Recovering the symmetry-breaking mechanism in the TI scenario}. We have advanced above that, starting from the Schrödinger operator $\mathcal{L}(t)$, a diagonal operator $S^{-1}(t)\mathcal{L}(t)S(t)$ can be constructed through an appropriate nonunitary transformation $S(t)$. Consequently, for a TI Hamiltonian $H$, it is straightforward to assume that a TI $S$ leads to the
diagonal form $S^{-1}\mathcal{L}(t)S=$ $S^{-1}HS - i\hbar\partial_{t}$ and to the solution of the Schrödinger equation. In fact, for a TI $S$, the eigenvalue Eq. (\ref{EEL-S}) reduces to $H\ket{\varepsilon_{n}}=\varepsilon_{n}\ket{\varepsilon_{n}}$ with the energy eigenvectors $\ket{\varepsilon_{n}}=S\ket{n}$ and eigenvalues $\varepsilon_{n} = \dot{\alpha}_n$. Similarly, Eq. (\ref{EELs}) reduces to $H\left(\mathcal{I}\ket{\varepsilon_{n}}\right)=\varepsilon^{\ast}_{n}\left(\mathcal{I}\ket{\varepsilon_{n}}\right)$ in the TI scenario with TI antilinear symmetries. The condition of unbroken symmetry given by Eq. (\ref{13}) then reduces to $\mathcal{I}\ket{\varepsilon_{n}} =\zeta_{n}\ket{\varepsilon_{n}}$ and, consequently, Eq. (\ref{14}) results in the well-known real spectrum: $\varepsilon _{n}^{\ast }=\varepsilon _{n}$. The SSB occurs when $H$ and $\mathcal{I}$ no longer share the same eigenstates, resulting in complex-conjugate eigenvalue pairs $\varepsilon_{n}$ and $\varepsilon^{\ast}_{n}$. Therefore, the method exposed above for TD $H(t)$ and $\mathcal{I}(t)$, with the central role played by the geometrical phases in the equality (\ref{14}), becomes clearer in light of the standard results for a TI $H$ and $\mathcal{I}$, where only the dynamical phases take place in the solution $\left\vert \psi
(t)\right\rangle $. If we had considered a nonunitary TD transformation $S(t)$, we could have obtained a SSB for autonomous Hamiltonians with the occurrence of geometrical phases in addition to dynamical ones.

\textit{The} $\mathcal{PT}$\textit{-symmetry breaking in the non-Hermitian dynamical Casimir effect}. The dynamical Casimir effect is the creation of photons from the quantum vacuum due to non-adiabatic changes of boundaries confining the electromagnetic field \cite{Moore}. In the non-Hermitian setting, it can be modeled as a $\mathcal{PT}$-symmetric, unbalanced parametric amplifier \cite{Cius}, described by the non-Hermitian Hamiltonian
\begin{equation}
H(t)=\omega (t)\left( a^{\dagger }a+\frac{1}{2}\right) +i\chi (t)\left(
\alpha a^{\dagger 2}-\beta a^{2}\right) ,  \label{15}
\end{equation}%
where the cavity frequency is modulated as $\omega (t)=\omega _{0}\left[ 1-\epsilon \cos \left( \kappa t\right)\right] $, with $\omega _{0}$ the cavity natural frequency, $\kappa $
the drive frequency of the moving boundary, and $\epsilon \ll 1$ a small modulation depth. The strength of the parametric amplification is $\chi (t)=\dot{\omega}(t)/4\omega (t)$, with $\alpha $ and $\beta $ being real parameters characterizing the unbalanced amplification such that $\mathcal{PT}%
H(t)\mathcal{PT}=H(-t)$, since under $\mathcal{PT}$: $a\rightarrow-a,a^{\dagger}\rightarrow-a^{\dagger},i\rightarrow-i$. Much work has been devoted \cite{FM2, Ponte, BeyondPT, Swanson} to non-Hermitian quadratic Hamiltonians of the form (\ref{15}) and it turns out that the simplest form for the metric operator 
\begin{equation}
\rho =\exp \left[ \frac{1}{2}\ln \left( \frac{\beta }{\alpha }\right) \left(
a^{\dagger }a+\frac{1}{2}\right) \right]  \label{16}
\end{equation}%
makes the Hamiltonian $H(t)$ pseudo-Hermitian without imposing constraints on its TD parameters. In addition, we recover the Law's Hermitian effective Hamiltonian \cite{Law} and the standard identity metric for $\alpha =\beta =1$.

In a convenient interaction picture defined by the operator $U(t)=\exp \left[i\xi (t)\left( a^{\dagger }a+1/2\right) \right] $ \cite{Viktor}, the Hamiltonian (\ref{15}) becomes 
\begin{equation}
V(t)=\Delta \left( a^{\dagger }a+\frac{1}{2}\right) +i\chi (t)\left( \alpha
a^{\dagger 2}e^{2i\xi (t)}-\beta a^{2}e^{-2i\xi (t)}\right) ,  \label{17}
\end{equation}
where we have defined the detuning $\Delta =\omega _{0}-\kappa /2\geq 0$ and
\begin{equation}
\xi (t)=\frac{\kappa t}{2}\left[ 1-\frac{2\omega _{0}\epsilon }{\kappa }%
\frac{\sin \left( \kappa t\right) }{\kappa t}\right] .  \label{18}
\end{equation}%
Assuming the non-adiabatic regime $\kappa t\gg 1$, we obtain to a good approximation $\xi (t)\approx \kappa t/2$, and under the Rotating-Wave Approximation (RWA) this leads to a TI interaction
\begin{equation}
V_{\text{RWA}}=\Delta \left( a^{\dagger }a+\frac{1}{2}\right) - g \left( \alpha a^{\dagger 2}+\beta a^{2}\right),  \label{19}
\end{equation}
with $g=\epsilon\kappa/8$ the effective parametric interaction. The condition $g\ll\kappa$ required by the RWA is completely consistent with the small modulation depth, i.e. with the condition that $\epsilon\ll1$. Consequently, the transformed Schrödinger operator $\mathcal{\tilde{L}}(t)=U(t)\mathcal{L}(t)U^{\dag }(t)=V_{\text{RWA}}-i\hbar \partial _{t}$ can be diagonalized by the TI generalized squeezing operator \cite{BeyondPT}
\begin{equation}
S =\exp \left[ \frac{r}{2}\left( \alpha a^{\dagger 2}-\beta
a^{2}\right) \right] ,  \label{20}
\end{equation}%
with squeezing strength
\begin{equation}
r=\frac{1}{2\sqrt{\alpha \beta }}\tanh ^{-1}\left(\frac{2g\sqrt{\alpha \beta }}{%
\Delta }\right).  \label{21}
\end{equation}

From the diagonalized $\mathcal{\tilde{L}}(t)$, we derive the TI eigenvalues of $\mathcal{L}(t)$, which are $\dot{\alpha}_{n}=\sqrt{\Delta
^{2}-4g^{2}\alpha \beta }\left( n+1/2\right) $, and their corresponding eigenvectors $\left\vert n,t\right\rangle =U^\dagger(t)S \left\vert
n\right\rangle $. The unbroken symmetry regime follows from $\Delta>2g\sqrt{\alpha\beta}$, where $\dot{\alpha}_{n}$ and the squeezing parameter $r$ are both real, with $\dot{\alpha}_n$ satisfying Eq. (\ref{14}). Thus, using $\mathcal{PT}\ket{n} = \zeta _{n} \ket{n}$ with $\zeta_n=\left(-1\right)^n$ \cite{PTn}, we obtain
\begin{equation}
\mathcal{PT}\ket{n,t} = \left(-1\right)^n\ket{n,-t},  \label{22}
\end{equation}
in accordance with the TD unbroken symmetry condition, Eq. (\ref{13}). Otherwise, for $\Delta<2g\sqrt{\alpha\beta}$, SSB occurs resulting in purely imaginary eigenvalues $\dot{\alpha}_{n}$ and squeezing parameter $r$. In this regime, Eq. (\ref{22}) is no longer valid. Fig. 1 illustrates the unbroken and broken regimes of $\mathcal{PT}$-symmetry in terms of the eigenvalues of $\mathcal{L}(t)$. At the exceptional point $\Delta=2g\sqrt{\alpha\beta}$, the eigenstates $\ket{n,t}$ and $\mathcal{PT}\ket{n,-t}$ of $\mathcal{L}(t)$ coalesce into a single eigenstate via Eq. (\ref{22}).

\begin{figure}[h]
	\centering
	\includegraphics[width=\textwidth,height=60mm,keepaspectratio]{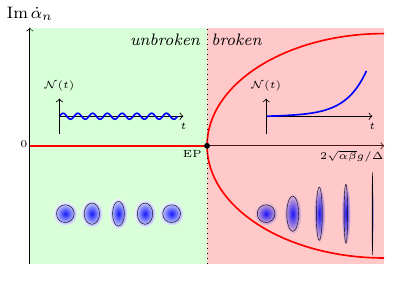}
	\caption{\textbf{Spontaneous symmetry breaking in the dynamical Casimir effect.} In the unbroken regime,  the eigenvalues $\dot{\alpha}_{n}(t)$ of the Schrödinger operator are real and the Casimir photons cannot be created since the number of photons $\mathcal{N}(t)$ oscillates with an amplitude less than unity. In the broken regime, the eigenvalues become complex-conjugate pairs, leading to a hyperbolic growth in the Casimir photons. The number of photons corresponds to states with large (small) squeezing in the field quadratures in the broken (unbroken) regime. The exceptional point (EP) occurs at $2\sqrt{\alpha\beta}g=\Delta$, marking the transition from unbroken to broken $\mathcal{PT}$-symmetry.}
\end{figure}

Next, from Eq. (\ref{SEsol}) we compute the average number of created Casimir photon (see SM), $\mathcal{N}(t)=\left\langle \psi (t)\left\vert a^{\dag
}a\right\vert \psi (t)\right\rangle _{\rho }$, which obeys $\ddot{\mathcal{N}}(t)+\Omega ^{2}\mathcal{N}(t)=8g^{2}\alpha \beta $ with $\Omega ^{2}=4\left( \Delta ^{2}-4g^{2}\alpha \beta \right) $. The general solution for $\mathcal{N}(t)$, starting from the vacuum state, is given by 
\begin{equation}
\mathcal{N}(t)=\frac{16g^{2}\alpha \beta }{\Omega ^{2}}\sin ^{2}\left( \frac{\Omega t}{2}\right) \text{.}  \label{23}
\end{equation}
For the unbroken regime (real $\Omega $), we obtain only the sinusoidal oscillatory solution with an amplitude less than unity. On the other hand, for the broken regime ($\Omega =2i\sqrt{4g^{2}\alpha \beta -\Delta ^{2}}$), we obtain a well-known hyperbolic growth of the Casimir photon number. Therefore, the dynamical Casimir effect, whether Hermitian ($\alpha=\beta=1$) or non-Hermitian ($\alpha\neq\beta$), is achieved only around the exceptional point, with broken $\mathcal{PT}$-symmetry resulting from the SSB. Otherwise, Casimir photons cannot be created in the unbroken regime.

Going beyond the RWA requires retaining the rapidly oscillating terms in $V(t)$, so that $V(t)=V_{\rm RWA}+\delta V(t)$, where $\delta V(t)$ is the explicitly TD correction neglected in Eq. (\ref{19}), given by
\begin{equation}
\delta V(t)=g\left(\alpha a^{\dagger2}e^{2i\kappa t}
+\beta a^2e^{-2i\kappa t}\right).
\end{equation}
Diagonalizing the transformed Schrödinger operator now requires a TD squeezing operator,
\begin{equation}
S(t)=\exp\left[ \frac{r(t)}{2}\left( \alpha a^{\dagger2}e^{i\theta(t)} -\beta a^2e^{-i\theta(t)} \right)\right],
\end{equation}
where $r(t)=r+\delta r(t)$ and $\theta(t)=\delta\theta(t)$ are perturbative
solutions around the RWA values, which are $\theta=0$ and $r$ as given in Eq.~(\ref{21}). To leading order, the corrections are
\begin{subequations}
	\begin{align}
		\delta r(t) & =\frac{g}{\kappa}\left[1-\cos(2\kappa t)\right], \\
		\delta\theta(t) & = -\frac{\Delta}{\kappa}\sin(2\kappa t).
	\end{align}
\end{subequations}
These corrections are perturbative in the small modulation since $g,\Delta\ll\kappa$. After accounting for corrections beyond the RWA, both eigenvectors and eigenvalues of $L(t)$ become explicitly TD: the eigenvectors are $\ket{n,t} = U^\dagger(t) S(t)\ket{n}$, whereas the corresponding TD eigenvalues are given by
\begin{equation}
\dot{\alpha}_n(t)\thickapprox
\left[
\frac{\Omega}{2}
+\left(\Delta-\frac{\Omega}{2}\right)\cos(2\kappa t)
\right]
\left(n+\frac12\right). \label{bRWA}
\end{equation}
This result shows that the TD correction neglected in the RWA does not change the exceptional point, $\Delta=2g\sqrt{\alpha\beta}$, but changes the way the symmetry condition is realized since the LR phases $\alpha_n(t)$ are no longer linear in time. For $\Delta>2g\sqrt{\alpha\beta}$, $\Omega$ is real, and $\alpha_n(t)$ obtained from Eq. (\ref{bRWA}) is an odd function of time that satisfies Eq. (\ref{14}). Moreover, because
$\delta\theta(t)$ and $\delta r(t)$ are odd and even functions of time, respectively, Eq. (\ref{13}) remains satisfied when $\mathcal{PT}$ is unbroken, both within and beyond the RWA. SSB emerges for $\Delta<2g\sqrt{\alpha\beta}$, when $\Omega$ becomes purely imaginary and yields imaginary contributions to the LR phases, causing the corresponding components of the system state to be amplified or attenuated over time.

\textit{Beyond $\mathcal{PT}$-symmetry.} A $\mathcal{PT}$-symmetric Hamiltonian may also exhibit an additional antilinear symmetry $\mathcal{I}(t)$ beyond the usual $\mathcal{PT}$. To construct such a symmetry, we introduce a TD linear operator $\mathcal{C}(t)$ based on the su(1,1) group algebra, so that $\mathcal{I}(t)=\mathcal{C}(t)\mathcal{PT}$, generalizing the standard $\mathcal{CPT}$-symmetry \cite{BBJ}. In order to determine $\mathcal{C}(t)$, we first note that  $I(t)=U^{-1}(t)Se^{i\pi\left(a^{\dagger}a+1/2\right)}S^{-1}U(t)$ is a nonlinear LR invariant \cite{MCM} associated with $H(t)$, sharing the same eigenvectors of $\mathcal{L}(t)$ with TI eigenvalues $\left(-1\right)^{n}$. From Eq. (\ref{IA}), we can deduce 
\begin{equation}
\mathcal{C}(t)=I(-t)=e^{i\phi\left(a^{\dagger}a+\frac{1}{2}\right)-i\mu\left(\alpha a^{\dagger2}e^{i\theta(t)}+\beta a^{2}e^{-i\theta(t)}\right)},
\end{equation}
where $\phi=\pi\cosh\left(2\sqrt{\alpha\beta}r\right)$, $\theta(t)=2\xi(t)$, and $\mu=-\pi\sinh\left(2\sqrt{\alpha\beta}r\right)/2\sqrt{\alpha\beta}$. Furthermore, it is easy to verify from Eq. (\ref{22}) that the condition for the unbroken regime is also satisfied for explicitly TD symmetries beyond $\mathcal{PT}$:
\begin{equation}
\mathcal{C}(t)\mathcal{PT}\ket{n,t} = + \ket{n,-t}.
\end{equation}
This result is in complete analogy with the TI scenario where the $\mathcal{CPT}$-symmetry is used as a positive-definite metric of the system with $\mathcal{C}$ acting as the measurement of the sign of the $\mathcal{PT}$ operator \cite{BBJ}. In the TD scenario, $\mathcal{CPT}$-symmetry also plays an important role in identifying positive-definite metrics for pseudo-Hermitian systems. In fact, the operator $\Xi(t)=\rho\mathcal{I}(t)$ defines an antilinear metric satisfying the TD pseudo-Hermiticity relation $i\hbar\dot{\Xi}(t)=\Xi(t)H(t)-H^{\dagger}(-t)\Xi(t)$, as proposed in \cite{BeyondPT}, which ensures the norm conservation.

In summary, while for autonomous systems the coalescence of the eigenenergies of the Hamiltonian indicates the phase transition from broken to unbroken symmetry, we have shown here that for nonautonomous Hermitian or non-Hermitian systems this central role is played by the eigenvalues of the Schrödinger operator $\mathcal{L}(t)$. The eigenvalue equation for $\mathcal{L}(t)$ has not only revealed the SSB, but has also allowed us to propose an alternative method to Lewis \& Riesenfeld for solving the Schrödinger equation for nonautonomous systems.

\begin{flushleft}
\textbf{{\Large Acknowledgements}}
\end{flushleft}

The authors acknowledge financial support from CAPES, CNPq, and FAPESP, Brazilian agencies.

\section{Supplemental Material}
\renewcommand{\theequation}{S\arabic{equation}}
\setcounter{equation}{0}

\subsection{The LR theorem for non-Hermitian Hamiltonians}

In order to provide a consistent extension of the LR invariant theorem for TD pseudo-Hermitian Hamiltonians, we start by defining the Dyson map $\eta(t)$, which transforms the Schrödinger equation $\left[H(t) - i\hbar \partial_{t}\right]\ket{\psi(t)}=0$ for a non-Hermitian $H(t)$, into the equation $\left[h(t) - i\hbar \partial_{t}\right]\ket{\phi(t)}=0$ for $h(t)$, the Hermitian counterpart of $H(t)$, given by
\begin{equation}
h(t)=h^{\dagger}(t)=\eta(t)H(t)\eta^{-1}(t)-i\hbar\eta(t)\partial_{t}\eta^{-1}(t),
\label{eq: CounPartH}
\end{equation}
where $\ket{\phi(t)}=\eta(t)\ket{\psi(t)}$. The Dyson map $\eta(t)$ is connected with the metric operator through $\rho(t)=\eta^{\dagger}(t)\eta(t)$, defining what is called the  \textit{time-dependent pseudo-Hermiticity relation}
\begin{equation}
H^{\dagger}(t)\rho(t)-\rho(t)H(t)=i\hbar\partial_{t}\rho(t). \label{S2}
\end{equation}

Considering a linear invariant operator $\Lambda(t)$ satisfying
\begin{equation}
i\hbar \partial_{t}\Lambda(t)+\left[ \Lambda(t),h(t)\right]=0,
\end{equation}
the LR theorem for $h(t)$ asserts that, given the eigenvalue equation $\Lambda(t)\ket{\varphi_{n}(t)}=\zeta_{n}\ket{\varphi_{n}(t)}$ with TI $\zeta_{n}$, the general solution of the Schrödinger equation is a superposition of the orthonormal eigenvectors of $\Lambda(t)$,
\begin{equation}
\left\vert \phi (t)\right\rangle =\sum_{n}c_{n}e^{-\frac{i}{\hbar }\alpha
_{n}(t)}\left\vert \varphi _{n}(t)\right\rangle , \label{eq: phi}
\end{equation}%
where the LR phases $\alpha _{n}(t)$ are determined by 
\begin{equation}
\dot{\alpha}_{n}(t)\delta _{n,m}=\left\langle \varphi _{n}(t)\left\vert
\left[ h(t)-i\hbar \partial _{t}\right]\right\vert \varphi _{n}(t)\right\rangle . \label{eq: LR-h}
\end{equation}%
Once the invariant $\Lambda(t)$ related to $h(t)$ is found, one can easily construct a dynamical invariant $I(t)$ related to the pseudo-Hermitian Hamiltonian $H(t)$ through the inverse transformation 
\begin{equation}
I(t)=\eta ^{-1}(t)\Lambda(t)\eta (t), \label{eq: ILambda}
\end{equation}%
where $I(t)$ satisfies the linear invariant condition 
\begin{equation}
i\hbar \partial_{t} I(t)+\left[ I(t),H(t)\right] =0.
\label{eq: IL}
\end{equation}%
The eigenvectors of $I(t)$, obtained from those of $\Lambda(t)$ by the inverse Dyson map 
\begin{equation}
\left\vert n,t\right\rangle =\eta
^{-1}(t)\left\vert \varphi _{n}(t)\right\rangle, \label{eq: IDM}
\end{equation}
 are orthonormal
with respect to the metric operator $\rho (t)$:
\begin{equation}
\braket{m,t|n,t}_{\rho (t)} = \left\langle m,t\left\vert \rho (t)\right\vert n,t\right\rangle =\delta
_{m,n}.  \label{eq: Ortho}
\end{equation}%
Furthermore, from Eqs. \eqref{eq: phi} and \eqref{eq: IDM}, the general solution of the Schrödinger equation for $H(t)$ becomes
\begin{equation}
\left\vert \psi (t)\right\rangle =\eta ^{-1}(t)\left\vert \phi
(t)\right\rangle =\sum_{n}c_{n}e^{-\frac{i}{\hbar }\alpha _{n}(t)}\left\vert
n,t\right\rangle,
\end{equation}%
where, by inserting Eq. \eqref{eq: CounPartH} into Eq. \eqref{eq: LR-h}, the LR phases are given by
\begin{equation}
\dot{\alpha}_{n}(t)\delta _{n,m}=\left\langle m,t\left\vert \left[H(t)-i\hbar
\partial _{t}\right]\right\vert n,t\right\rangle _{\rho (t)}.  \label{eq: LRphase}
\end{equation}

Instead of starting from the transformation in Eq. \eqref{eq: ILambda}, connecting both invariants for the Hermitian and non-Hermitian Hamiltonians, the authors in Ref. \citep{LR-NH} have assumed Eqs.  \eqref{eq: IL} and \eqref{eq: Ortho} from the beginning in order to generalize the LR theorem. As a final observation, we note that when a non-Hermitian $H(t)$ reduces to a Hermitian operator, the Dyson map becomes a unitary transformation, such that $\eta^{\dagger}(t)=\eta^{-1}(t)$, and consequently the metric becomes the identity operator: $\rho(t)=\eta^{\dagger}(t)\eta(t)=\mathbbm{1}$.

\subsection{Extending the Wigner theorem for pseudo-unitary transformations}

For a pseudo-Hermitian Hamiltonian $H(t)$ satisfying Eq. (\ref{S2}), a TD transformation $S(t)$ maps the Schrödinger equation into $i\hbar\partial_t\ket{\psi'(t)}=V\ket{\psi'(t)}$, where $\ket{\psi'(t)}=S^{-1}(t)\ket{\psi(t)}$  and 
\begin{equation}
V(t)=S^{-1}(t)H(t)S(t)+i\hbar\frac{\partial S^{-1}(t)}{\partial t}S(t).
\end{equation}
Considering the state $\ket{\psi(t)}$ defined in A, the norm conservation imposes
\begin{equation}
\braket{\psi'\vert \psi'}_\rho =  \braket{\psi'|\rho|\psi'}=\braket{\psi|\psi}_\rho =1
\end{equation}
leading to the pseudo-unitary relation
\begin{equation}
S^{\dagger}(t) \rho(t) S(t) = \rho(t).
\end{equation}
For Hermitian Hamiltonians, where $\rho=\mathbbm{1}$, we recover the well-known Wigner theorem stating that all transformations of quantum states must be represented by unitary operator, $S^{\dagger}S=\mathbbm{1}$.

\subsection{Antilinear symmetry condition in terms of $\mathcal{L}(t)$}
To obtain Eq. (\ref{IAL}), we act the operator (\ref{IA}), representing the TD antilinear symmetry condition, on an arbitrary state $\ket{\cdot}$ such that
\begin{equation*}
\left[H(-t)+i\hbar\partial_{t}\right]\mathcal{I}(t)\ket{\cdot} -\mathcal{I}(t)\left[H(t)-i\hbar\partial_{t}\right]\ket{\cdot} =0.
\end{equation*}
With the Schrödinger operator $\mathcal{L}(t)=H(t)-i\hbar\partial_{t}$, this becomes 
\begin{equation*}
\left[\mathcal{L}(-t)\mathcal{I}(t)-\mathcal{I}(t)\mathcal{L}(t)\right]\ket{\cdot}=0,
\end{equation*}
and, since $\mathcal{I}(t)$ is invertible, we conclude
\begin{equation}
\mathcal{I}(t)\mathcal{L}(t)\mathcal{I}^{-1}(t)=\mathcal{L}(-t).
\end{equation}

\subsection{Casimir photons and squeezed states}

The pseudo-Hermitian quadratures of the field, $X_{1}$ and $X_{2}$, are determined by the Dyson map $\eta=\rho^{1/2}$, from the Hermitian quadratures $x_{1}=(a+a^{\dagger})/2$ and $x_{2}=(a-a^{\dagger})/2i$ by using the inverse transformation $X=\eta^{-1}x\eta$. From Eq. (\ref{16}), we obtain 
\begin{subequations}
	\begin{align}
			X_{1} & = \left(ae^{\frac{1}{4}\ln\left(\frac{\beta}{\alpha}\right)}+a^{\dagger}e^{\frac{1}{4}\ln\left(\frac{\alpha}{\beta}\right)}\right)/2, \\
			X_{2} & = \left(ae^{\frac{1}{4}\ln\left(\frac{\beta}{\alpha}\right)}-a^{\dagger}e^{\frac{1}{4}\ln\left(\frac{\alpha}{\beta}\right)}\right)/2i.
	\end{align}
\end{subequations}
Starting from the vacuum state, we easily see that the expectation values of $X_{1}$ and $X_{2}$ remain always null. Furthermore, the expectation values $\left\langle X_{1}^{2}\right\rangle _{\rho}$,  $\left\langle X_{2}^{2}\right\rangle _{\rho}$, and $\mathcal{N}(t)=\left\langle a^{\dagger}a\right\rangle _{\rho}$ are obtained by the following coupled equations
\begin{subequations}
	\label{S16}
	\begin{align}
		i\frac{d\mathcal{N}}{dt} & = 2g\left(\beta\mathcal{A}-\alpha\mathcal{B}\right), \\ i\frac{d\mathcal{A}}{dt} & = 2\Delta\mathcal{A}-4\alpha g\left(\mathcal{N}+1/2\right), \\ i\frac{d\mathcal{B}}{dt} & = -2\Delta\mathcal{B}+4\beta g\left(\mathcal{N}+1/2\right),
	\end{align}
\end{subequations}
where we have defined 
\begin{subequations}
	\begin{align}
		\braket{ X_{1}^{2}}_\rho & = \left(1+\mathcal{N}\right)/2+\left(\alpha\mathcal{B}e^{+i\kappa t}+\beta\mathcal{A}e^{-i\kappa t}\right)/4\sqrt{\alpha\beta}, \\
		\braket{ X_{2}^{2}}_\rho & = \left(1+\mathcal{N}\right)/2-\left(\alpha\mathcal{B}e^{+i\kappa t}+\beta\mathcal{A}e^{-i\kappa t}\right)/4\sqrt{\alpha\beta}.
	\end{align}
\end{subequations}
The system (\ref{S16}) can be easily solved by taking into account the constant of motion $\mathcal{N}(t)=g\left(\alpha\mathcal{B}(t)+\beta\mathcal{A}(t)\right)/\Delta $ under the initial conditions $\ensuremath{\mathcal{A}(0)}=\ensuremath{\mathcal{B}(0)=0}$ and $\mathcal{N}(0)=0$. We thus obtain
\begin{subequations}
	\begin{align}
		\mathcal{A}(t)	& =\frac{2\alpha g}{\Omega}\left\{ \frac{2\Delta}{\Omega}\left[1-\cos\left(\Omega t\right)\right]+i\sin\left(\Omega t\right)\right\} \\ 
		\mathcal{B}(t) & = \frac{2\beta g}{\Omega}\left\{ \frac{2\Delta}{\Omega}\left[1-\cos\left(\Omega t\right)\right]-i\sin\left(\Omega t\right)\right\} \\
		\mathcal{N}(t) & = \frac{16\alpha\beta g^{2}}{\Omega^{2}}\sin^{2}\left(\Omega t/2\right),
	\end{align}
\end{subequations}
where $\Omega ^{2}=4\left( \Delta ^{2}-4g^{2}\alpha \beta \right)$.

The uncertainties in the quadratures of the field are given by $\Delta X_{1}=\sqrt{\left\langle X_{1}^{2}\right\rangle _{\rho}-\left\langle X_{1}\right\rangle _{\rho}^{2}}$ and $\Delta X_{2}=\sqrt{\left\langle X_{2}^{2}\right\rangle _{\rho}-\left\langle X_{2}\right\rangle _{\rho}^{2}}$. The system evolves from the vacuum to a squeezed vacuum state, which can be characterized through the rotated quadratures $Y_{1}=X_{1}\cos\left(\varphi/2\right)+X_{2}\sin\left(\varphi/2\right)$ and $Y_{2}=-X_{1}\sin\left(\varphi/2\right)+X_{2}\cos\left(\varphi/2\right)$ along a direction defined by $\varphi(t)=\pi/2-\kappa t$. In the unbroken regime, the uncertainties are given by $\Delta Y_{1} \thickapprox 1/2+\left(\sqrt{\alpha\beta}g/4\Delta\right)\sin\left(\Delta t/2\right)$ and $\Delta Y_{2} \thickapprox 1/2-\left(\sqrt{\alpha\beta}g/4\Delta\right)\sin\left(\Delta t/2\right)$, both varying slightly around the vacuum fluctuations. However, when the SSB takes place, the squeezing degree $s=2\sqrt{\alpha\beta}gt$ increases monotonically with time such that $\Delta Y_{1}=e^{s}/2$ and $\Delta Y_{2}=e^{-s}/2.$


\begin{thebibliography}{99}
\bibitem{PTn} Starting from $a^{\dagger}a\ket{n}=n\ket{n}$, it is straightforward to derive $a^{\dagger}a\left(\mathcal{PT}\ket{n}\right)=n\left(\mathcal{PT}\ket{n} \right)$ and, consequently, $\mathcal{PT}\ket{n} =\zeta_{n}\ket{n}$ with $\left|\zeta_{n}\right|^{2}=1$. When we redefine the Fock state $\ket{0}$  by adding a phase $\zeta_{0}^{-1/2}$, we obtain $\mathcal{PT}\ket{0}=\ket{0}$  and therefore the eigenvalue equation $\mathcal{PT}\ket{n} = \left(-1\right)^n \ket{n}$.
\end{thebibliography}
\end{document}